# Origin of time reversal symmetry breaking in $Y_{1-y}Ca_yBa_2Cu_3O_{7-x}$.

Yoram Dagan and Guy Deutscher.

*School of Physics and Astronomy, Raymond and Beverly Sackler Faculty of Exact Sciences, 69978 Tel Aviv Israel*

Abstract

We have studied the Zero Bias Conductance Peak (ZBCP) of the tunneling conductance measured on (1,1,0) oriented $Y_{1-y}Ca_yBa_2Cu_3O_{7-x}$ thin films as a function of doping and of magnetic field. A spontaneous (zero field) split of the ZBCP was observed only in overdoped samples (either by O or by Ca). The magnitude of this split was found to be linear in doping. All samples exhibited a magnetic field splitting, also strongly doping dependent. The field susceptibility $\chi = \frac{d\delta}{dH}\bigg|_{H\rightarrow 0}$ diverges at the point at which spontaneous ZBCP splitting occurs, its inverse value, $\chi^{-1}$, following a linear doping dependence on both the underdoped and overdoped sides. We discuss these results in terms of recent theoretical models of Time Reversal Symmetry Breaking (TRSB).

It is by now well established that the Order Parameter (OP) in the High $T_c$ Cuprates has a dominant *d*-wave component.[1] A consequence of such a symmetry is that low energy surface bound states are created at all surfaces whose normal is not aligned with the (1,0,0) or equivalent direction.[2] These states are due to interference effects between quasiparticles that undergo Andreev reflections from lobes of the OP having opposite signs. If the OP has the pure $d_{x^2-y^2}$ symmetry, zero energy states are created. When probed through a tunnel junction with a normal electrode, these states give a ZBCP.[3] In the presence of surface currents the degeneracy between states carrying currents of opposite signs is lifted, and the ZBCP is split by energy $2\delta$, proportional to the amplitude of the surface currents.[4] Field induced splitting was studied in detail in recent publications.[5], [6], [7], [8] A spontaneous split of the ZBCP interpreted as resulting from spontaneous surface currents, or equivalently to spontaneous TRSB was also reported in Refs.[6], [7], [8] .Such currents imply the existence of an *is* or a *$id_{xy}$*, component, whose amplitude is equal, at low temperatures, to the value of the zero field splitting $\delta(0)$.[9] However, in other tunneling experiments on $YBa_2Cu_3O_{7-x}$ spontaneous ZBCP splitting was not observed.[10], [11], [12], [13], yet it appears to be systematic in Ca overdoped $Y_{1-y}Ca_yBa_2Cu_3O_{7-x}$ [14], [15].

In this contribution we resolve this apparent experimental irreproducibility. We show that the spontaneous splitting of the ZBCP is doping dependent, being always present beyond a certain doping level (near the optimum, which is the doping level giving the maximum $T_c$), and never at lower ones. The amplitude of $\delta(0)$ varies as the distance from optimum doping, (p-$p_c$) where p is the hole concentration and $p_c$ its value at the optimum level. In addition, we find that at that the field splitting is in fact also strongly doping dependent, the inverse susceptibility $\chi^{-1}$ varies as $|p-p_c|$. The doping dependences of $\delta(0)$ and of $\chi$ are in good agreement with recent theoretical models for a Quantum Critical Point (QCP), where the symmetry changes from pure $d_{x^2-y^2}$ to $d_{x^2-y^2}+id_{xy}$.

(1,1,0) oriented $YBa_2Cu_3O_{7-x}$ films of various doping levels near the optimal doping were grown on (1,1,0) $SrTiO_3$ substrates, using RF and DC off-axis sputtering. A $PrBa_2Cu_3O_{7-x}$ template was used in order to reduce the (1,0,3) orientations, following the method used by Poelders *et al.*[16]. Garcia Lopez *et al.*[17] showed that changing the oxygen pressure and the amount of atomic and ionic oxygen in the plasma could control the amount of oxygen loaded in the $YBa_2Cu_3O_{7-x}$ films during the growth. In addition, they showed that the film remains oxidized if quenched at room temperature. We have used this method of Garcia Lopez *et al.* to grow the slightly oxygen overdoped samples, with the doping level controlled by the amount of water vapor added to the plasma. This affects the amount of atomic oxygen in the plasma.[18] By increasing it, we obtained a reduction of $T_c$ and of , the position of the *d*-wave gap feature. Underdoping was achieved by annealing in a low oxygen pressure. Additionally we have prepared overdoped $Y_{0.9}Ca_{0.1}Ba_2Cu_3O_{7-x}$ and $Y_{0.8}Ca_{0.2}Ba_2Cu_3O_{7-x}$ films as described elsewhere.[15], [19].

To make sure that we have indeed obtained overdoped films we have annealed them in a low oxygen pressure environment at $650°C$, a $T_c$ of 90K was then retrieved as expected. In addition we have found the R(T) characteristics to be different in overdoped and underdoped samples. In fig.1 the R(T) characteristics of one overdoped and one underdoped $YBa_2Cu_3O_{7-x}$ films are presented. The contacts were put in a square with its sides parallel to the [0,0,1] and to the [1,$\bar{1}$,0] directions. The film resistances are measured with the current running parallel to the ab planes, the in plane resistance presented in the figure is calculated using the procedure of ref.[20]. One can note that R(T) of the overdoped sample has the expected positive curvature [21] while the underdoped film exhibits the downwards deviation from linearity, signaling the pseudogap temperature, [22] at $T > T_c$.

We have made a systematic study of the spontaneous and field induced ZBCP splitting in 14 YBa$_2$Cu$_3$O$_{7-x}$ samples, with thickness ranging from 2400Å to 800Å, as a function of the level of oxygen doping. The most underdoped films had T$_c$=83.6K and the most overdoped films had T$_c$=85.6K, down-set the transition width was about 1K in the overdoped samples and 2K in the underdoped ones. We have also studied 2 Y$_{0.9}$Ca$_{0.1}$Ba$_2$Cu$_3$O$_{7-x}$ and 2 Y$_{0.8}$Ca$_{0.2}$Ba$_2$Cu$_3$O$_{7-x}$ samples. The Ca doped films had a wide resistive transition (with a width of 5K). The counter electrode used for junction fabrication was Indium. Junction preparation and characterization are described elsewhere.[15], [11], [23]. All measurements were taken at T=4.2K. Field splitting of the ZBCP was observed in all samples when the magnetic field was applied parallel to the surface of the film, along the c direction. Upon applying the field parallel to the surface of the film, but along the [1,$\bar{1}$,0] direction only a weak splitting, one or two orders of magnitude smaller was observed, as already reported.[7], [8] The *d*-wave gap feature, however, is sensitive to a field applied in both directions, and may be even somewhat more sensitive to a field applied parallel to the CuO planes.

Two typical data sets obtained on YBa$_2$Cu$_3$O$_{7-x}$ films are shown Fig.2 (underdoped, T$_c$=85.9K) and Fig.3 (overdoped, T$_c$=85.9K down-set). Insets show how the ZBCP splits as a function of magnetic field. In both cases, the low field behavior of δ(H) is linear, the slope determining the susceptibility, χ. A spontaneous splitting of the ZBCP is seen only the in overdoped sample (δ(0)=2.4mV). Notice that the conductance peak at about 16mV, due to the *d*-wave gap, is well defined in both samples.

In fig.4 we present the conductance characteristics at various magnetic fields for an (overdoped) Y$_{0.9}$Ca$_{0.1}$Ba$_2$Cu$_3$O$_{7-x}$ sample. We note that the *d*-wave gap feature is here wider than in the YBa$_2$Cu$_3$O$_{7-x}$ samples, presumably due to the disorder introduced by Ca doping.[24] A zero field splitting δ(0)=2.4mV is seen. In the inset we show the field splitting versus magnetic field. The characteristics of $Y_{0.8}Ca_{0.2}Ba_2Cu_3O_{7-x}$ samples are qualitatively similar to those already shown in refs. [15], [19].

The data shown Fig.2 to 4 is typical, in the sense that spontaneous splitting at 4.2K is never observed in underdoped samples, but always in overdoped ones (either by O or by Ca). This confirms the findings of refs. [14], [15]. Field splitting exists in both underdoped and overdoped samples, but is doping dependent. In fact, as shown earlier, at optimum doping $\delta(H)$ is not linear, but rather varies as $\sqrt{H}$.[11]

Fig.5 summarizes our results ($YBa_2Cu_3O_{7-x}$ samples) for the doping dependence of $\delta(0)$ and $\chi^{-1}$. $\delta(0)$ and $\chi^{-1}$ are plotted against $(\Delta_{max}-\Delta)^{1/2}$, where $\Delta_{max}$ is the value of the *d*-wave gap at optimum doping, and $\Delta$, its value for the sample under consideration, as measured on each junction [23]. As shown in the insert of fig.5, $\Delta/kT_c$ is constant within our experimental accuracy, in the range of doping investigated here. Since it is known that $T_c$ has a parabolic doping dependence near optimum doping, $(\Delta_{max}-\Delta)^{1/2}$ is proportional to $|p-p_c|$ where p is the hole doping level and $p_c$ its value at optimum doping.[25]. As seen in Fig.5, $\delta(0) \propto (p-p_c)$ and $\chi^{-1} \propto |p-p_c|$. We have found these results to be thickness independent. [26]

We shall now discuss theoretical models dealing with TRSB in the High $T_c$ cuprates.

Fogelström *et al.*[9] have attributed the zero field ZBCP splitting to surface currents induced by spontaneous *is* surface component, and the field splitting to Meissner currents. However, having found the field splitting to be independent of film thickness, we must conclude that it cannot be principally due to Meissner currents.[26] Tanuma *et al.*[27] have calculated the doping dependence of the amplitude of a possible spontaneous surface *is* component. Their results do show a weakening in the underdoped regime, but very little dependence around optimum doping.

Another possible origin of spontaneous surface currents is a modification of the bulk OP, in the form of an $id_{xy}$ component. Laughlin has shown that such a component is energetically favorable in the presence of a field B, applied perpendicular to the CuO planes. [28] This is because it involves a magnetic moment (while an $is$ component does not). Deutscher et al.[15] have proposed that both the spontaneous and the field induced splitting are due to a bulk $id_{xy}$ component, whose value is the OP of an $d_{x^2-y^2}$ to $d_{x^2-y^2}+id_{xy}$ phase transition. This OP minimizes the free energy, which is proposed to have the form:

$$F = a\delta^2 + b|\delta|^3 + cB\delta \qquad (1)$$

Where $a \propto (p_c-p)$ and b, c (c<0) are given by Laughlin.[28] Minimization of the free energy with respect to $\delta$ gives at $p>p_c$ the equilibrium value:

$$\delta = \frac{-a + \sqrt{a^2 - 3bcB}}{3b} \qquad (2)$$

This model predicts that $\chi$ diverges at the critical point $p=p_c$ as it is approached from both sides. It has already been shown that $\chi$ is very large at optimum doping, while the field splitting being in fact nonlinear in field, varying rather as $\sqrt{H}$.[11] This is in agreement with Eq.2, assuming that the critical point is at, or very near, optimum doping. The results, that we report here, show that indeed $\chi^{-1} \propto |p-p_c|^\gamma$ with $\gamma=1\pm0.05$ for the underdoped range and $\gamma=1\pm0.3$ for the overdoped range, $\delta(0) \propto (p-p_c)^\beta$ with $\beta=1\pm0.3$, in agreement with the model with $\gamma$ and $\beta$ being the critical indices for the susceptibility and of the OP. The obvious advantage of the model is that it allows a complete description of the data, based on a single free energy expression. Simmilar results can be obtained with the critical point being slightly on the overdoped side. If one assumes $\gamma=\beta=1$, the best fit to the data is obtained for a critical point at $[\Delta_{max}-\Delta]^{1/2}=0.16$ $(mV)^{1/2}$, on the overdoped side. Indications for a critical point on the overdoped side have been given by Bernhard et al..[29]

Recently, Vojta *et al.*[30] have also conjectured, on the basis of angle resolved scattering rate measurements by Vala *et al.*[31], that there exists in the vicinity of the superconducting phase of the cuprates, a QCP where a $d_{x^2-y^2}$ to a $d_{x^2-y^2}+id_{xy}$ transition takes place.

Khveshchenko and Paaske [32] have calculated the free energy near a $d_{x^2-y^2}$ to a $d_{x^2-y^2}+id_{xy}$ QCP, taking into account the interactions between the quasiparticles near the nodes and fluctuations of the OP. They obtained $\beta=0.87$ and $\gamma\approx1$, also in agreement with our data.

Assuming a doping driven phase transition allows a complete description of the data, particularly of the fact that the susceptibility diverges at the doping level beyond which spontaneous ZBCP splitting exist. We are therefore inclined to believe that it is the most likely explanation of the spontaneous and field induced symmetry breaking effects that we have reported here.

One of us (G.D.) is indebted to Carlo di Castro Claudio Castellani and Marco Grilli for an in-depth discussion of our data, and for pointing out that the critical indices calculated by Khveshchenko and Paaske imply γ of the order of 1 also in agreement with our experiments. We are indebted to Y. Tanaka and A. Kohen for useful discussions. This work was supported in part by the Heinrich Hertz - Minerva Center for High Temperature Superconductivity, by the Israeli Science Foundation and by the Oren Family Chair of Experimental Solid State Physics.

**Correspondence and requests for materials should be addressed to Y. Dagan (e-mail: dovratb@tau.ac.il).**

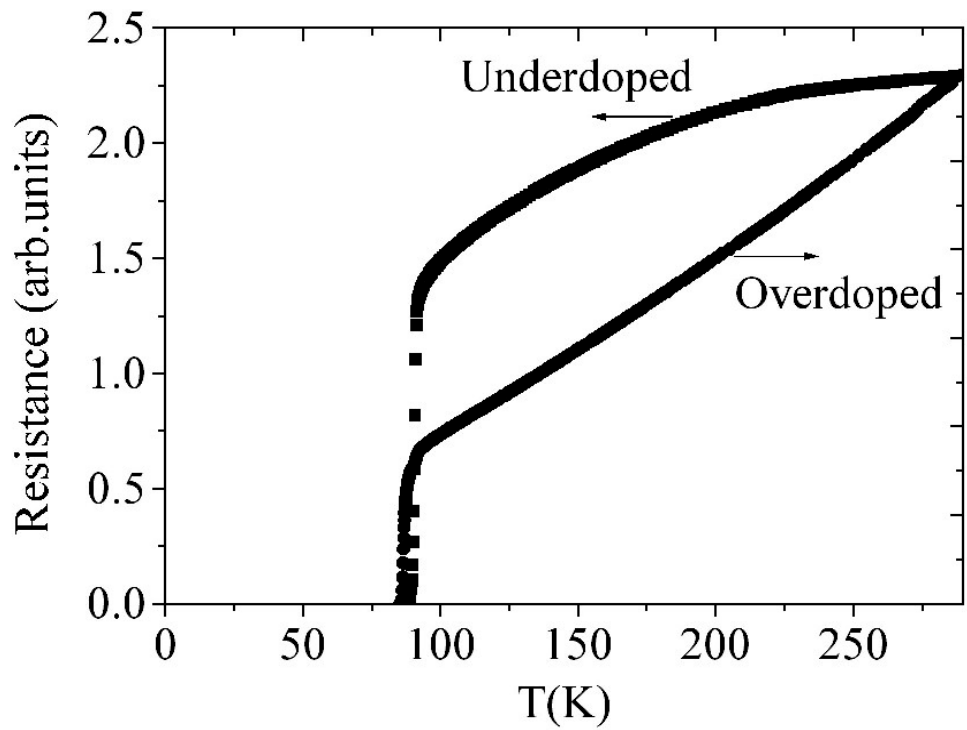

**Figure 1** The in-plane resistance versus temperature for overdoped and underdoped $Y_1Ba_2Cu_3O_{7-x}$ samples. We note the positive curvature of the resistance in the overdoped case, and the downward deviation from linearity in the underdoped one.

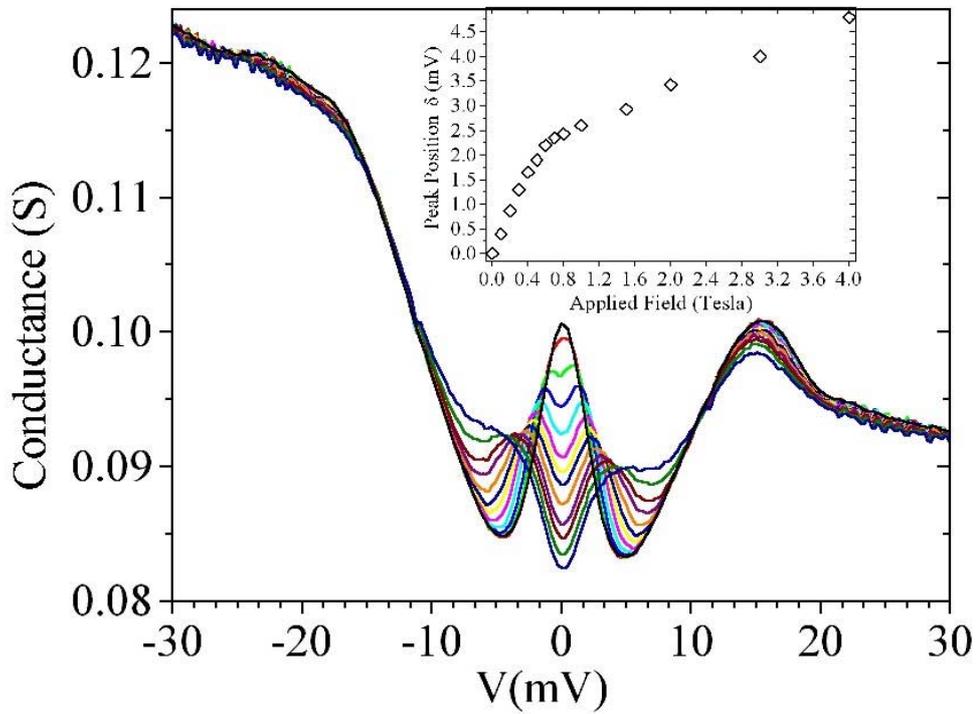

**Figure 2** The conductance versus voltage for a slightly underdoped $Y_1Ba_2Cu_3O_{7-x}$ sample ($T_c$=88K down-set) at various magnetic fields: 0T (black), 0.1T (red), 0.2T (green), 0.3T (blue), 0.4T (cyan), 0.5T (magenta), 0.6T (yellow), 0.7T (navy), 1T (orange), 1.5T (purple), 2T (wine), 3T (olive), 4T (royal). **Insert:** The split peak position (on the positive bias side) versus magnetic field for the sample.

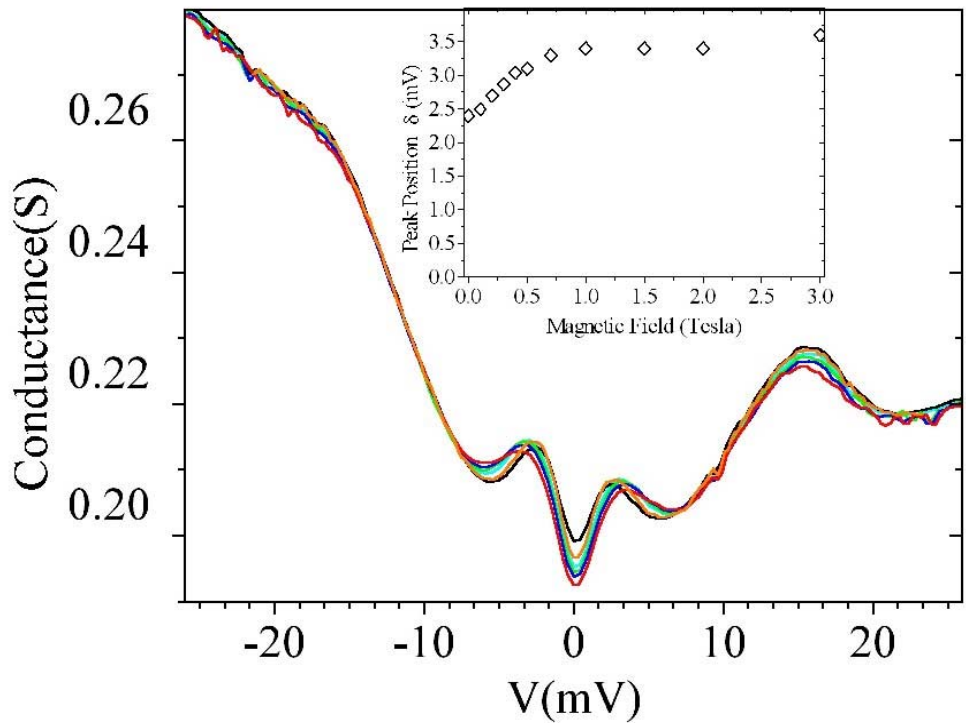

**Figure 3** The conductance versus voltage for a slightly overdoped $Y_1Ba_2Cu_3O_{7-x}$ sample ($T_c$=85.9K down-set) at various magnetic fields: 0T (black), 0.2T (orange), 0.5T (cyan), 1T (green), 3T (blue), 6T (red). **Insert:** The split peak position (on the positive bias side) versus magnetic field for the sample.

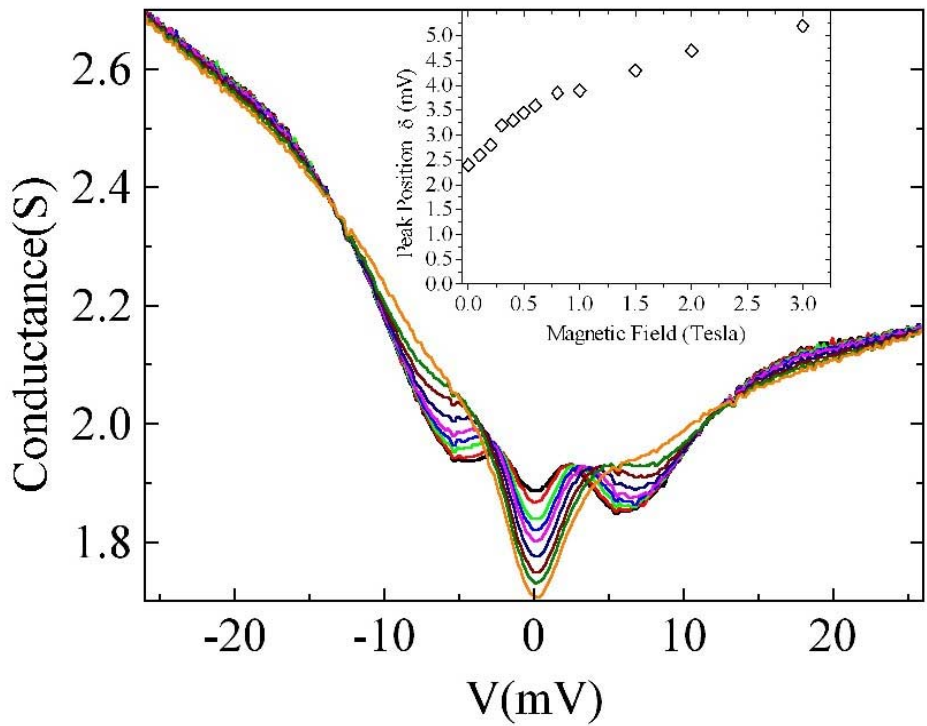

**Figure 4** The conductance versus voltage for a $Y_{0.9}Ca_{0.1}Ba_2Cu_3O_{7-x}$ sample ($T_c$=67.5K down-set) at various magnetic fields: 0T (black), 0.1T (red), 0.2T (green), 0.3T (blue), 0.5T (magenta), 1T (navy), 2T (wine), 3T (olive), 6T (orange). **Insert:** The split peak position (on the positive bias side) versus magnetic field for the sample.

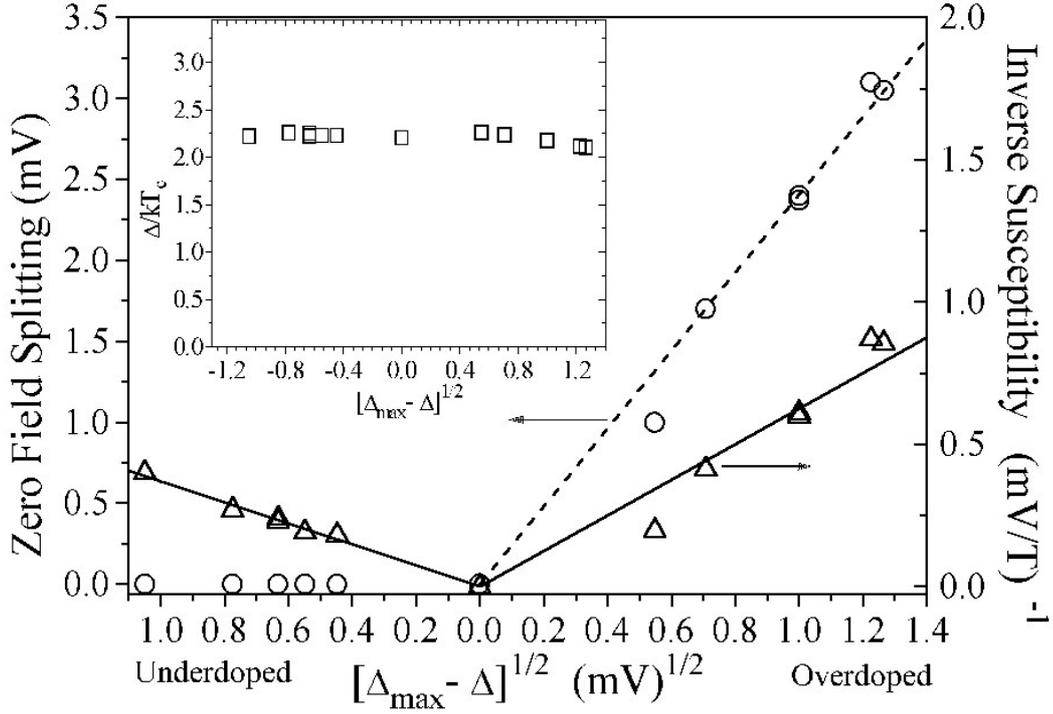

**Figure 5** The spontaneous (at zero field) splitting of the zero bias conductance peak versus $[\Delta_{max}-\Delta]^{1/2}$ (circles) for doping ranging from slightly underdoped ($T_c$=83.6K down set) to slightly overdoped ($T_c$=85.6K down set) . $[\Delta_{max}-\Delta]^{1/2}$ is a quantity proportional to the doping level (see text).Triangles: the inverse susceptibility $\chi^{-1}$ for the same samples. The upper bound of $\chi^{-1}$ for the sample with $(\Delta_{max}-\Delta)$=0 is 0.08[mV/T]$^{-1}$. Solid lines: linear fits for both the underdoped and overdoped ranges. The lines extrapolate to zero at the same doping level where the spontaneous splitting appears. Dashed line linear fit for $\delta(0)$ on the overdoped side. Inset: $2\Delta/kT_{cW}$ for the samples measured.